# Optical second harmonic generation from Wannier excitons

Thomas G. Pedersen[1] and Horia D. Cornean[2]

[1]Department of Physics and Nanotechnology, Aalborg University, DK-9220 Aalborg Øst, Denmark

[2]Department of Mathematical Sciences, Aalborg University, DK-9220 Aalborg Øst, Denmark

Excitonic effects in the linear optical response of semiconductors are well-known and the subject of countless experimental and theoretical studies. For the technologically important second order nonlinear response, however, description of excitonic effects has proved to be difficult. In this work, a simplified three-band Wannier exciton model of cubic semiconductors is applied and a *closed form* expression for the complex second harmonic response function including broadening is derived. Our calculated spectra are found to be in excellent agreement with the measured response near the band edge. In addition, a very substantial enhancement of the nonlinear response is predicted for the transparency region.

Second order nonlinearities are of great technological importance in e.g. frequency conversion and electro-optics. Zinc-blende and Wurtzite semiconductors such as GaAs and GaN are prominent examples of materials with large optical nonlinearities. For these crystal classes, several theoretical attempts to describe the nonlinear response based on first principles have been published [1-4]. Reasonable agreement with experimental second harmonic generation (SHG) spectra has been obtained in the case of GaAs and GaP [1-3]. In comparison to calculations of linear optical properties, however, the level of sophistication in still low. A crucial ingredient in an accurate theory of optical nonlinearity is excitonic effects. To the knowledge of the present authors, the works by Chang *et al.* [4] and Leitsmann *et al.* [5] are the only systematic attempts at including excitons effects in the SHG response of semiconductors. These works used the *ab initio* Bethe-Salpeter approach, which very succesfully incorporates excitonic effects in the linear optical response [6,7]. There are severe limitations to this method in the nonlinear case, however. Thus, the approach of Ref. [4] is restricted to the low-frequency response and the reported SHG spectra only covers the photon energy region well below the fundamental energy gap. The method used in Ref. [5] does not suffer from this restriction but computational cost limits the density of *k*-points and e.g. discrete excitonic resonances near band edges are not resolved. Experimental SHG spectra [8,9] provide firm evidence of profound excitonic effects, however, in particular near band edge transitions. Thus, there is a clear need for alternative approaches to the nonlinear excitonic response.

For the linear optical response, decisive insight has been gained from studies of the Wannier exciton limit. The seminal work by Elliott [10] provided an explicit expression (Elliott formula) for the optical absorption due to bound and continuum excitons in terms of the effective exciton Rydberg *Ry*. This work was subsequently generalized by Tanguy [11] to cover both real and imaginary parts of the optical response and to include line broadening. The basic premises of these calculations are that (i) two parabolic bands are considered, and (ii) transition momentum matrix elements between single-particle states are constant throughout the Brillouin zone. These approximations are excellent for excitations in the vicinity of the band gap. In the present work, the Wannier exciton model is used to calculate a "nonlinear Elliott formula" for the optical SHG response. We consider cubic semiconductors and derive an explicit expression for the complex second order susceptibility including broadening. Exciton effects are shown to produce substantial changes in the spectra, in particular around band edge transitions. In the limit of vanishing Coulomb interaction, our spectra agree reasonably with previous full-band calculations. We compare calculations with and without excitons to measured SHG spectra for ZnS [9]. Excellent agreement is found for the Coulombic theory whereas the free-carrier result completely fails to describe the data. Also, a comparison is made with experimental GaAs data near the $E_1$ transition [12], which shows that excitonic effects improve the agreement between calculated and measured SHG spectra. Importantly, our theory provides a prediction for the excitonic enhancement of the low-frequency SHG response. A rather substantial enhancement ranging from 25% to 200% is predicted for some important materials.

The second order nonlinear response of a semiconductor necessarily involves interband transitions. In the present calculation we restrict attention to direct-gap zinc-blende structures as one of the most important groups of nonlinear semiconductors. Hence, a minimal model [9] corresponds to the simplified three-band structure shown in Fig. 1. In this model, a single occupied valence band ($\Gamma_{15v} \equiv v$) and two empty conduction bands ($\Gamma_{1c} \equiv c$ and $\Gamma_{15c} \equiv \tilde{c}$) are the only bands considered. The $v \to c$ and $c \to \tilde{c}$ energy gaps are denoted $E_g$ and $\Delta$, respectively. In general, intraband transitions are known to contribute to the nonlinear response but such terms cannot easily be included in a Wannier exciton model and they are consequently neglected. Our starting point, therefore, is the pure interband susceptibility given by e.g. Eq.(B1) of Ref.[2] or Eq.(2) of Ref. [3], that is, we ignore the additional terms $\eta$ and $\sigma$ [2] representing intraband modulation of the linear susceptibility and interband polarization effects on the intraband motion, respectively. In Ref. [5], Leitsmann *et al.* demonstrate that all three terms may be combined into compact two- and three-band expressions. Hence, the combined three-band expression could be applied in order to partly include excitonic intraband effects. In the present work, focus is on resonant interband transitions, however, and we consequently ignore all intraband related effects. In the notation of Ref. [5], this is equivalent to ignoring two-band terms as well as intraband-derived three band terms. In cubic semiconductors, the second order nonlinear response is solely due to the *xyz* element of the nonlinear susceptibility $\vec{\chi}^{(2)}(-2\omega;\omega,\omega)$. In the spin-degenerate three-band case, after inclusion of broadening and conversion of position to momentum matrix elements, this element is given by

$$\chi^{(2)}_{xyz}(-2\omega;\omega,\omega) = \frac{2ie^3\hbar^3}{V\varepsilon_0 m_0^3} \sum_{\substack{n\in vc \\ m\in v\tilde{c}}} \frac{P^x_{0n}P^y_{nm}P^z_{m0}}{E_{n0}E_{mn}E_{m0}} \{F(\Omega_2,\Omega_1) + F(-\Omega_1,-\Omega_2) \\ + F(-\Omega_1,\Omega_1) + F(-\Omega_2,-\Omega_1) + F(\Omega_1,\Omega_2) + F(\Omega_1,-\Omega_1)\},$$
(1)

where $V$ is the crystal volume, $F(a,b) = (E_{n0} - a)^{-1}(E_{m0} - b)^{-1}$ and $\Omega_1 = \hbar\omega + i\hbar\Gamma$ and $\Omega_2 = 2\hbar\omega + i\hbar\Gamma$ with $\Gamma$ a phenomenological line broadening. Moreover, index 0 refers to the ground state and indices $n$ and $m$ refer to exciton states composed of single-excitations from the valence band ($v$) to the low ($c$) and high ($\tilde{c}$) conduction band, respectively. The transition energies $E_{ij}$ and momentum matrix elements $\vec{P}_{ij}$ must be obtained from the Wannier exciton eigenstates. A particular eigenstate such as $|n\rangle$ is a linear superposition in $\vec{k}$-space $|n\rangle = \sum_{\vec{k}} \tilde{\psi}_n(\vec{k}) |v\vec{k} \to c\vec{k}\rangle$ of single-particle excitations $|v\vec{k} \to c\vec{k}\rangle$ from valence band $v$ to conduction band $c$. We neglect band-mixing introduced by the Coulomb interaction, i.e. mixed excitons containing both $|v\vec{k} \to c\vec{k}\rangle$ and $|v\vec{k} \to \tilde{c}\vec{k}\rangle$ contributions. When the $\vec{k}$-dependence of the single-particle momentum matrix elements $\vec{p}_{vc}$ is neglected, an exciton momentum matrix element such as $\vec{P}_{0n}$ is given by

$$\vec{P}_{0n} = \vec{p}_{vc} \sum_{\vec{k}} \tilde{\psi}_n(\vec{k}) = \sqrt{V} \vec{p}_{vc} \psi_n(0),$$
(2)

where $\psi_n(\vec{r})$ is the inverse Fourier transform of $\tilde{\psi}_n(\vec{k})$. In a similar manner, it is found that

$$\vec{P}_{nm} = \vec{p}_{c\tilde{c}} \sum_{\vec{k}} \tilde{\psi}_n^*(\vec{k}) \tilde{\psi}_m(\vec{k}) = \vec{p}_{c\tilde{c}} \int \psi_n^*(\vec{r}) \psi_m(\vec{r}) d^3 r. \tag{3}$$

The Fourier transforms $\psi_n(\vec{r})$ and $\psi_m(\vec{r})$ are solutions to separate Wannier equations since their respective conduction bands are characterized by effective masses $m_e$ and $\tilde{m}_e$, c.f. Fig. 1. Here, however, we shall ignore this complication and take the $\Gamma_{1c}$ effective mass $m_e$ to characterize both conduction bands. With this simplification, Eq.(3) reduces to $\vec{P}_{nm} = \vec{p}_{c\tilde{c}} \delta_{nm}$ and, furthermore, the exciton energies are related via $E_{m0} = E_{n0} + \Delta \equiv \tilde{E}_{n0}$. The susceptibility then reduces to the much simpler form

$$\begin{aligned}\chi_{xyz}^{(2)}(-2\omega;\omega,\omega) = \chi_0^{(2)} \{ S(\Omega_2,\Omega_1) + S(-\Omega_1,-\Omega_2) \\ + S(-\Omega_1,\Omega_1) + S(-\Omega_2,-\Omega_1) + S(\Omega_1,\Omega_2) + S(\Omega_1,-\Omega_1) \},\end{aligned} \tag{4}$$

with

$$\chi_0^{(2)} = i \frac{2^{1/2} 3 e^3 \mu^{3/2}}{\pi \varepsilon_0 m_0^3 E_0^{5/2} \Delta} p_{vc}^x p_{c\tilde{c}}^y p_{\tilde{c}v}^z, \tag{5}$$

where $\mu = m_e m_h / (m_e + m_h)$ is the reduced mass and $E_0^{5/2} \equiv E_g^{1/2} \tilde{E}_g^{1/2} (E_g^{1/2} + \tilde{E}_g^{1/2})^3$ with $\tilde{E}_g \equiv E_g + \Delta$. In fact, $\chi_0^{(2)}$ is the low-frequency limit $\chi_{xyz}^{(2)}(0;0,0)$ in the case of vanishing Coulomb effects $Ry \to 0$. It is a real-valued constant because the triple momentum product at the zone centre is purely imaginary [3]. The remaining dimensionless summations are of the form

$$S(a,b) = \frac{4}{3}\pi a_0^3 Ry^{3/2} E_0^{5/2} \sum_n \frac{|\psi_n(0)|^2}{E_{n0} \tilde{E}_{n0} (E_{n0}-a)(\tilde{E}_{n0}-b)}. \qquad (6)$$

Here, $a_0$ is the exciton Bohr radius and just as in the linear case, this result shows that only $s$-excitons contribute to the response. Using explicit expressions for these eigenstates [10], the sum in Eq.(6) can be evaluated. Both bound and continuum excitons contribute and for the latter part the sum can be converted into the following integral:

$$S_{cont}(a,b) = \frac{16}{3}\pi^2 Ry^{3/2} E_0^{5/2} \int_0^\infty \frac{x^5 dx}{(1-e^{-x})(\alpha^2 + E_g x^2)(\alpha^2 + \tilde{E}_g x^2)(\alpha^2 + (E_g-a)x^2)(\alpha^2 + (\tilde{E}_g-b)x^2)}, \qquad (7)$$

where $\alpha^2 = 4\pi^2 Ry$. In a manner similar to the linear case [11], this integral is split into partial fractions and the final result is

$$S_{cont}(a,b) = -\frac{Ry^{1/2} E_0^{5/2}}{3}\left\{\frac{1}{a\Delta(\Delta-b)}g\left(\sqrt{\frac{Ry}{E_g}}\right) + \frac{1}{b\Delta(\Delta+a)}g\left(\sqrt{\frac{Ry}{\tilde{E}_g}}\right)\right. \\ \left. -\frac{1}{a(\Delta+a)(\Delta+a-b)}g\left(\sqrt{\frac{Ry}{E_g-a}}\right) - \frac{1}{b(\Delta-b)(\Delta+a-b)}g\left(\sqrt{\frac{Ry}{\tilde{E}_g-b}}\right)\right\} \qquad (8)$$

with $g(z) = 2\psi(z) + 2\ln(z) + 1/z$, where $\psi$ is the digamma function. In the low frequency limit, a simple expression for $S_{cont}(0,0)$ can be given in terms of $g(z)$ and $g'(z) = 2\psi'(z) + 2/z - 1/z^2$.

The effect of excitons on the nonlinear susceptibility can be illustrated for a number of important materials using the parameters listed in Table 1.

| Material | $E_g$ [eV] | $\Delta$ [eV] | $Ry$ [meV] |
|----------|------------|---------------|------------|
| GaAs | 1.52 | 3.2 | 4.2 |
| ZnSe | 2.7 | 5.1 | 19 |
| ZnS | 3.7 | ~4.65 | 36 |
| CuCl | 3.4 | ~5.8 | 190 |

Table 1. Relevant material parameters for selected zinc-blende semiconductors [13].

The selected zinc-blende semiconductors range from cases of small (GaAs) over medium (ZnSe and ZnS) to large (CuCl) Coulomb effects as indicated by the effective Rydberg $Ry$. In Figs. 2 and 3, theoretical SHG spectra for GaAs and ZnS (using data from Table 1 and $\hbar\Gamma = 30$ meV) are compared with and without including excitons. The excitonic effect is obviously quite dramatic in ZnS but even in the low-$Ry$ case of GaAs a pronounced modification is observed. It is noted, moreover, that our free-carrier spectra (neglecting excitons) are quite similar in shape to the full-band calculations of e.g. Ref. [3] apart from obvious contributions from other bands and high-symmetry points at energies above the fundamental energy gap.

A crucial effect of Coulomb interactions is the shift of oscillator strength towards lower energies in the spectra. Hence, in particular in the case of ZnS, the fundamental band edge transition ($2\hbar\omega \sim E_g$) is significantly intensified by the accumulation of oscillator strength, as illustrated in

Fig. 3. In the inset, experimental SHG data from Ref. [9] is compared to the present theory. To simulate energetically remote contributions from higher bands and other parts of the Brillouin zone, a real-valued constant has been added to the theoretical response. This parameter as well as the pre-factor $\chi_0^{(2)}$ have been adjusted to fit the measured behavior below the band edge. In addition, the numerical value of the band gap is taken as 3.7 eV + $Ry$ because the experimental value (3.7 eV) determined from reflectivity measurements is not corrected for the exciton binding energy. Our calculated response including excitons is clearly seen to agree with the experimentally observed SHG response near the ZnS band edge. In comparison, the free-carrier model severely underestimates the response at the band edge. The response near the experimentally observed discrete 1S exciton resonance agrees with the Wannier exciton model but clearly not with the free-carrier calculation. For GaAs, to our knowledge, no experimental SHG data exist for two-photon energies in the vicinity of the fundamental gap $2\hbar\omega \approx E_g$ and so the calculated spectra of Fig. 2 cannot be directly compared to measurements. However, Bergfelt and Daum [12] have recorded SHG spectra around the characteristic $E_1$ transition with a room-temperature band gap $E_g^{(1)} \approx 2.9$ eV [14]. Hence, as a simplistic comparison with GaAs experiments we will apply the present theory to the $E_1$ transition using $E_g^{(1)}$ as the fundamental gap and taking $\Delta^{(1)} \approx 3$ eV [14] and $Ry = 4.2$ meV (for lack of a better estimate) for this transition. The corresponding spectra and the comparison with experimental data (adding, again, a constant high-energy contribution) are shown in Fig. 4. As in the case of ZnS, excitonic effects tend to improve agreement between theory and experiment, in particular for the distinct feature at resonance $\hbar\omega = E_g^{(1)}/2$. Clearly, the sharpness of the resonance is rather sensitive to excitonic effects even for materials with a relatively small excitonic binding energy such as GaAs.

A related consequence of the excitonic effect is the enhancement of the nonlinear response in the transparency region $\hbar\omega < E_g/2$, which is the important region for most applications. In the present framework, this Coulomb enhancement factor can be quantitatively estimated in the low-frequency limit by the value $\chi^{(2)}_{xyz}(0;0,0)/\chi^{(2)}_0 = 6S(0,0)$ ignoring energetically remote contributions. The Coulomb enhancement predicted by the Wannier exciton model is illustrated in Fig. 5 for the four selected materials. In addition, the continuous curve is an estimate using averaged values of $E_g$ and $\Delta$ from Table 1 and taking the effective Rydberg $Ry$ as a parameter. This curve quite closely follows a power law $\chi^{(2)}_{xyz}(0;0,0)/\chi^{(2)}_0 \approx 1 + 0.071 \cdot (Ry\,[\text{meV}])^{0.59}$. In fact, the power law explains the remarkable Coulomb enhancement even in low-$Ry$ materials such as GaAs for which the present model predicts an increase of roughly 25%. Our findings for the Coulomb enhancement in low-$Ry$ cases are in good agreement with the Bethe-Salpeter approach [4]. Clearly, excitonic enhancement is an important factor in the search for highly nonlinear materials. Moreover, an even larger increase can be expected in low-dimensional structures such as quantum wells and wires due to the enhanced binding of quantum confined excitons. Finally, is should be noted that an identical enhancement of the linear electro-optic response $\chi^{(2)}_{xyz}(-\omega;\omega,0)$ is predicted since the two nonlinear response functions become equal in the low-frequency limit.

In summary, the optical second harmonic response from Wannier excitons in cubic semiconductors is calculated analytically. A compact, closed form expression for the complex susceptibility including broadening is derived and applied to several important cases. The agreement with the measured excitonic nonlinear response near the fundamental band edge of ZnS is excellent. For the $E_1$ transition of GaAs, an enhanced band-edge response is predicted in agreement with experiments. The Coulomb enhancement of the nonlinearity in the transparency region is found to scale as a power law with the effective exciton Rydberg. As a result, a substantial increase of the nonlinear response ranging from 25% to 200% is predicted for several important materials.

Dr. Hans-Peter Wagner and Dr. Winfried Daum are gratefully acknowledged for providing the experimental data.

Figure 1. Minimal three-band model containing a single valence band and two conduction bands.

Figure 2 (color online). Real (upper panel) and imaginary (lower panel) SHG response of GaAs. Blue and red curves show calculations with and without excitons, respectively.

Figure 3 (color online). SHG response of ZnS. Inset: zoom of the band edge region including experimental data from Ref. [9].

Figure 4 (color online). Real (upper panel) and imaginary (lower panel) SHG response of GaAs near the $E_1$ transition. The inset shows a comparison of the absolute value of the response with experiment from Ref. [12].

Figure 5 (color online). Calculated Coulomb enhancement of the low-frequency nonlinear response. The squares are obtained using the parameters of Table 1. The curve is calculated from averaged band parameters taking the effective Rydberg $Ry$ as a continuous parameter.

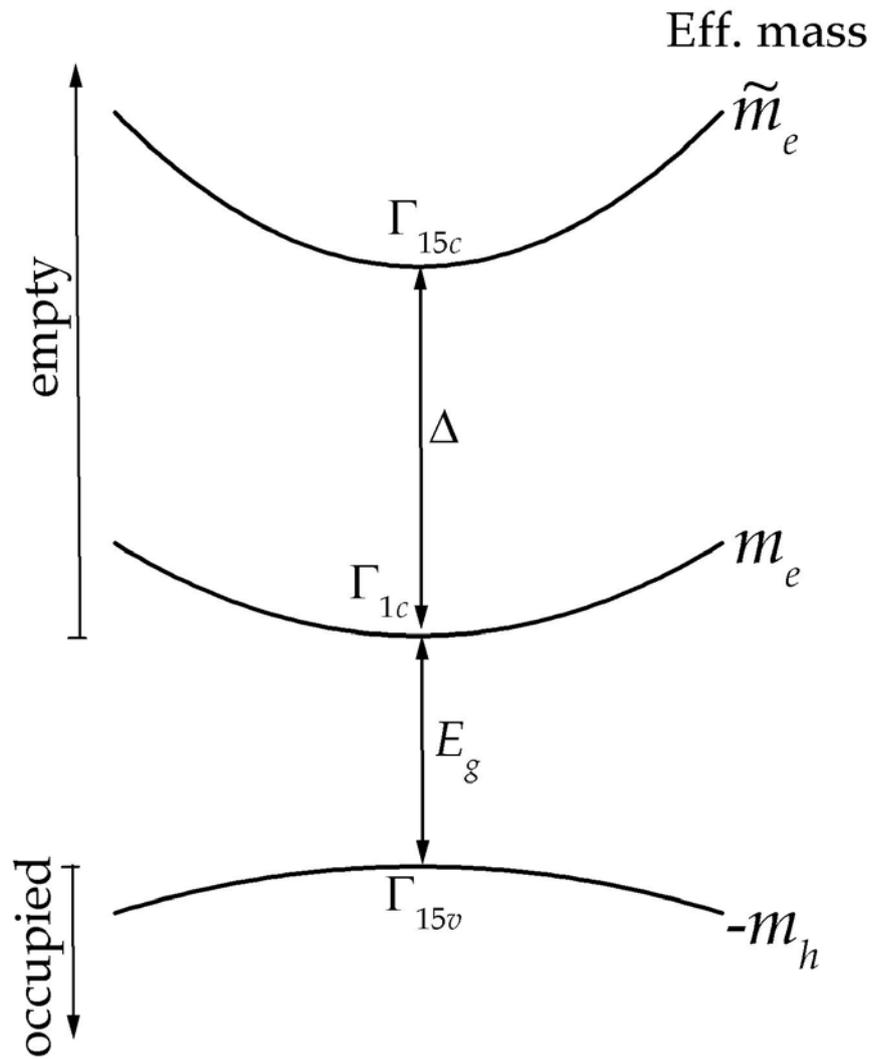

Fig. 1 T.G. Pedersen *et al.*

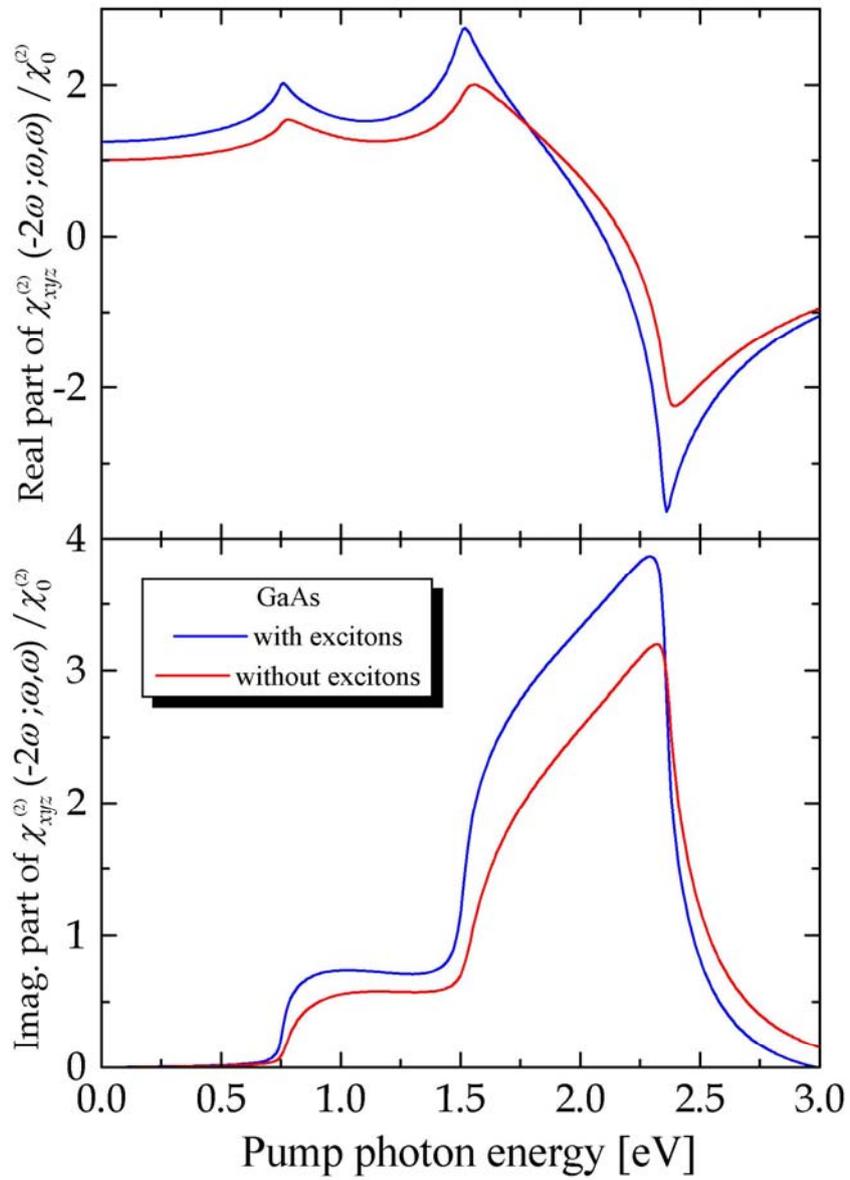

Fig. 2 T.G. Pedersen *et al.*

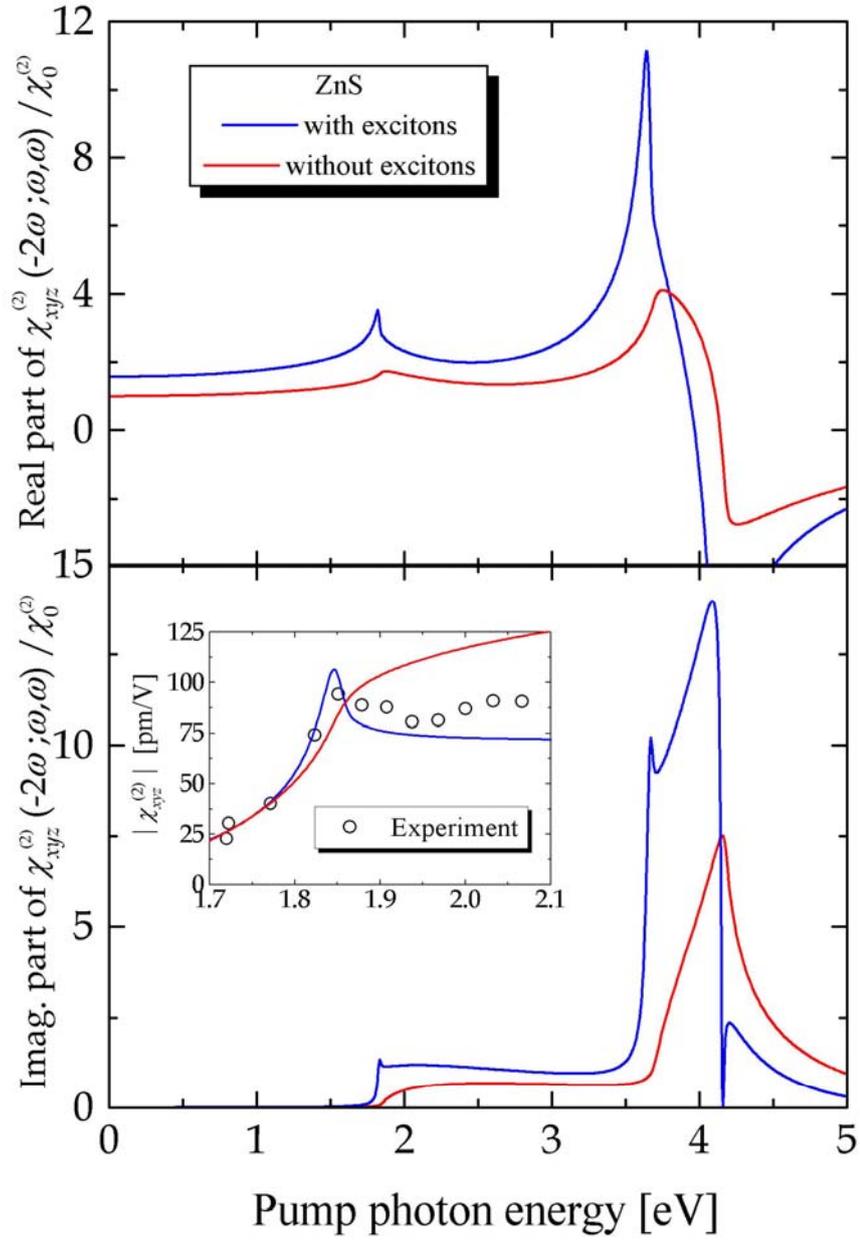

Fig. 3 T.G. Pedersen *et al.*

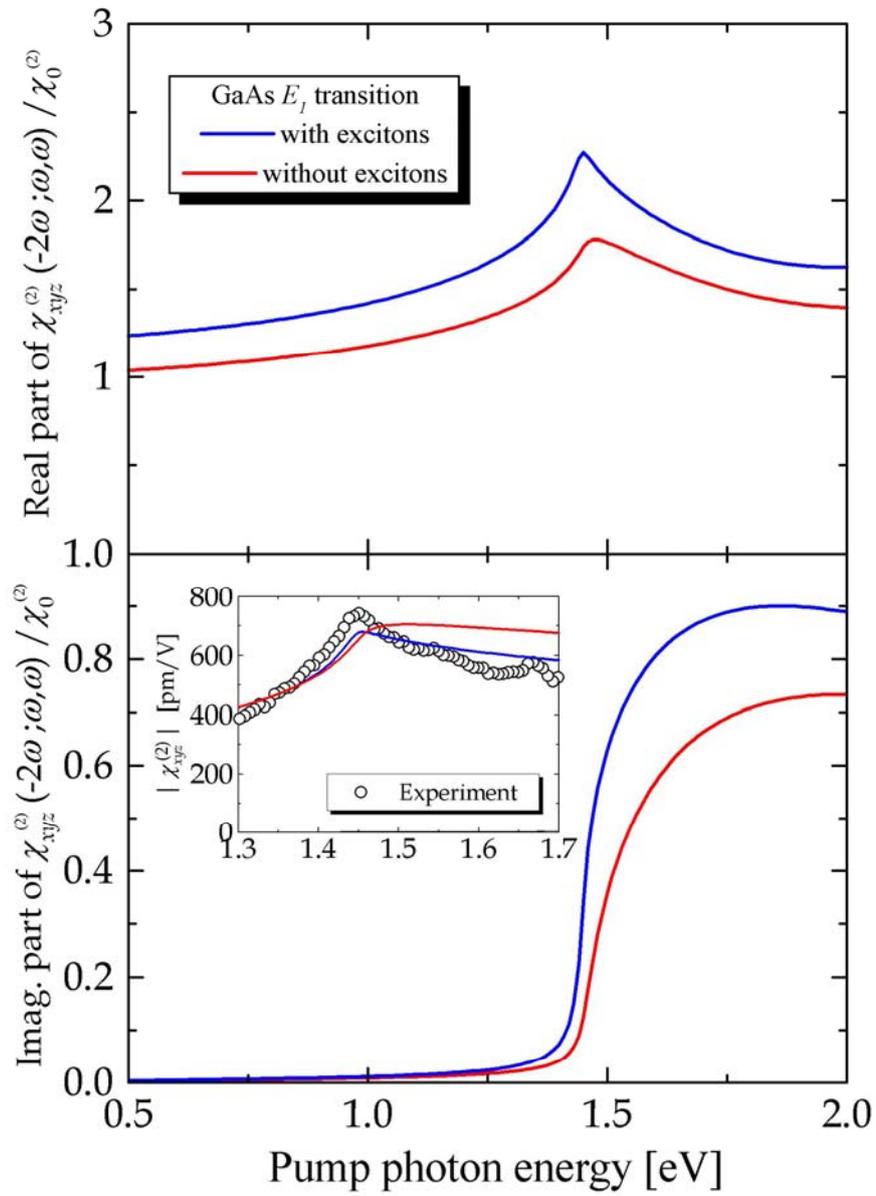

Fig. 4 T.G. Pedersen *et al.*

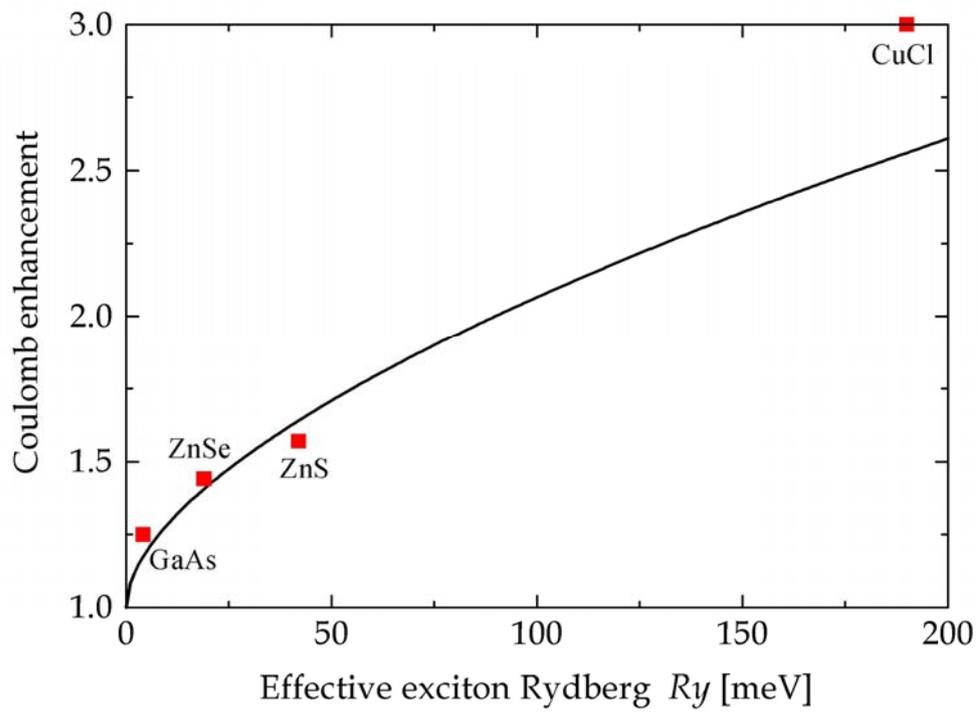

Fig. 5 T.G. Pedersen *et al.*